\documentstyle[aps,preprint]{revtex}
\begin{document}
\setcounter{page}{0}
\def\footnoterule{\kern-3pt \hrule width\hsize \kern3pt}
\tighten
\title{Gauged Nonlinear Sigma Model and\\
Boundary Diffeomorphism Algebra
}
\author{ Phillial  Oh\footnote{
Email address: ploh@dirac.skku.ac.kr
}}
\address{
Department of Physics and 
Basic Science Research Institute\\
Sung Kyun Kwan University, Suwon 440-746,
Korea }


\maketitle

\begin{abstract}
We  consider Chern-Simons gauged nonlinear sigma model  with boundary  
which has a  manifest bulk diffeomorphism invariance. We find that the Gauss's 
law can be solved explicitly when the nonlinear  sigma model is defined on
the Hermitian symmetric  space, and the original bulk theory completely
reduces to a boundary nonlinear sigma model with the target space of Hermitian
symmetric space. We also study the symplectic structure, compute the
diffeomorphism algebra on the boundary, and  find an (enlarged) Virasoro
algebra with classical central term.   
\end{abstract}

\vspace*{\fill}

\section{Introduction}

The Chern-Simons (CS) gravity theory \cite{ref:Ach} with boundary
\cite{btz} has  attracted a lot of recent attention  in  relation with black hole 
physics\cite{ref:Car}. In particular, it has been observed \cite{strom} that
the Virasoro algebra which lives on the boundary and carries classical central
charge \cite{henn,ban,carll} may provide  important clues in understanding the
microscopic origin of black hole entropy.

It is well known that  in the Chern-Simons theory\cite{ref:Des},
the Gauss constraint can be solved explicitly by the pure gauge condition, and
the bulk theory completely reduces to a boundary 
chiral WZW model on $G$ \cite{Bal}. If one imposes some extra
condition on the group element, reduction from the target space of $G$
to that of $G/H$ occurs \cite{pere}. For example, if one imposes an extra
condition  like $g^2=1$, $CP(N)$ nonlinear sigma model  (NLSM) can be obtained.
However, such process is ad hoc and  the reduced theory on $CP(N)$ has some
properties which cannot be obtained from the original WZW theory by simply
substituting $g^2=1$  (The enlarged Virasoro algebra of eq. (\ref{symmetric2}) is one 
such example.).  The purpose of this Letter is to invent some scheme
where the reduction to $G/H$ occurs directly as a consequence of Gauss' law
constraint. Using the coadjoint orbit method for NLSM \cite{oh},
we  introduce a model in which the Chern-Simons gauge field and NLSM on $G/H$
have topological and gauge invariant interaction.  
Gauss's law can be solved explicitly when the NLSM is
defined on the Hermitian symmetric spaces, and the original bulk theory
completely reduces to a boundary NLSM with the target space of
 Hermitian symmetric space. We also study the symplectic
structure of the boundary theory and compute the
diffeomorphism algebra by the standard Noether procedure. We
find that the diffeomorphism algebra becomes  the Virasoro algebra with 
classical central term if a suitable boundary condition is satisfied. We also
discover  an enlarged Virasoro algebra with elements of symmetric tensor
product.

\section{Symplectic Reduction and Boundary NLSM}
To describe the coupling of CS theory with NLSM, we introduce a 
coadjoint orbit variable $Q =gKg^{-1}$ on $G/H$, where $g\in G$ and 
$K\in {\cal H}$ is the
centralizer for the Lie algebra ${\cal H}$. 
Then, we consider  the Chern-Simons theory coupled with
NLSM in a diffeomorphism invariant manner on the disc ${D}$
\begin{equation}
L=\Gamma \int _{D} d^2 x \epsilon^{\mu \nu \rho}
\left<[Q, D_\mu Q] F_{\nu\rho}\right>+
\frac{\kappa}{2 \pi} \int _{D} d^2 x \epsilon^{\mu \nu \rho} \left< A_{\mu} 
\partial_{\nu} A_{\rho} +\frac{2}{3} A_{\mu} A_{\nu} A_{\rho}
\right>,
\label{eq:1}
\end{equation}
where $\left<\cdots\right>$ denotes trace and $D_\mu=\partial_\mu
+A_\mu$. Since we are interested in a complete reduction to the boundary
theory, we neglect the metric dependent part of the Lagrangian for the
variable $Q$  and consider only the manifest diffeomorphism-invariant theory. 
 The above Lagrangian also has the gauge  invariance  
\begin{equation}
g\rightarrow h g,~ Q\rightarrow h Q h^{ -1},
~ A\rightarrow h A h^{ -1}+h d h^{ -1},
~~~~(h\in G),
\end{equation}
and the constant coefficient $\Gamma (\equiv \kappa \alpha/2\pi)$ is not
quantized. It will be related with the solution of the Gauss's law constraint.
The Lagrangian  also has a local symmetry
$g\rightarrow gh~(h\in H), Q\rightarrow Q,~ A\rightarrow A$,
which makes the field $Q$  to take values in the homogeneous space $G/H$.
Up to a boundary term, (\ref{eq:1}) 
can be put into the following canonical form 
\begin{equation}
L=\frac{\kappa}{4 \pi}\int_{D} d^2 x \epsilon^{ij}
 \left ((A_i^a +2\alpha [Q, D_i Q] )\dot{A}^{a}_{j}
+\alpha [Q, F_{ij}]^a\dot Q^a
-A_0^a G^a_{ij}\right).
\label{eq:2}
\end{equation}
(Here, $\epsilon^{012} \equiv \epsilon^{12}\equiv 
1,~Q=Q^at^a,~ A_i =A^a_i t^a,~ F_{ij}=F_{ij}^a t^a,~F_{ij}^a=\partial_i A^a_j 
-\partial_j A^a _i +f^{abc} A_i^b A_j^c$, and the group generators $t^a$ 
satisfy $[t^a, t^b]=f^{abc} t^c,~\left<t^a t^b\right>=-\frac{1}{2}
\delta^{ab}$.) $G^a_{ij}$ is the Gauss' law constraint given by
\begin{equation}
G^a_{ij}=F_{ij}^a+2\alpha[D_iQ, D_jQ]^a-
2\alpha[Q,[Q,F_{ij}]]^a.
\label{gau}
\end{equation}

We shall take (\ref{eq:2}) as our starting point. 
Variation with respect to 
$A_0^a$ gives the Gauss' law constraint $G^a_{ij}=0$. 
We adopt the Hamiltonian symplectic method \cite{ref:Fad} and  solve the
constraint from the beginning. 
It turns out that the above constraint can be solved explicitly
when $Q$ corresponds to Hermitian symmetric spaces\cite{ford,oh}
with the  following Ansatz;
\begin{equation}
A_i^a=c[Q,~\partial_i Q]^a,
\label{eq:3}
\end{equation}
where $c$ is a constant to be determined.  
 Substituting into (\ref{gau})  and 
using the identities $[Q,[Q,\partial_i Q]]=-\partial_i Q,~
[Q, [\partial_i Q, \partial_j Q]]=0$ which are satisfied \cite{oh}
on Hermitian symmetric spaces,  we find that the constraint can be solved by
choosing 
\begin{equation}
c=-1\mp\sqrt{1/(1+2\alpha)}. ~~~~(\alpha>-\frac{1}{2})
\label{eq:4}
\end{equation}
When $\alpha=0$, $c$ can be
either $0$ or $-2$. In the latter case, $A_i^a=
-2[Q,~\partial_i Q]^a$
satisfies the zero curvature condition which in $SU(2)$ case has been studied
before in the context of vacuum structure of pure Yang-Mills theory\cite{jevi}
and in integrable models \cite{oh}.
 Substitution into the Lagrangian (\ref{eq:2})
gives 
\begin{equation} 
L=\gamma\oint  d\varphi <\partial_\varphi Q \dot Q>,
\label{lagrangian}
\end{equation}
where $\varphi$ denotes the angular coordinate on the boundary $\partial {D}$
of disc $D$ and $\gamma=
\kappa c (2\pi)^{-1}(c+2\alpha(1+c))$. Hence, the bulk theory has completely
reduced to boundary theory upon substitution of the solution of the Gauss' law
constraint \cite{namb}.  

In order to compute the symplectic structure, 
we follow the approach of
Ref. \cite{ohpark} to rewrite the Lagrangian in terms of the current defined
by $J_\varphi^a= \gamma[Q,~\partial_\varphi Q]^a$. 
Using the identities $[Q,[Q,\partial_ \varphi Q]]=-\partial_ \varphi Q$, 
we obtain
\begin{equation}
L=\oint d \varphi <J_\varphi [Q,\dot Q]>.
\label{synp1}
\end{equation}
which suggests the following canonical 1-form
\begin{equation}
\theta =\oint d \varphi <J_\varphi [Q, dQ]>.
\label{symp1}
\end{equation}
Then, using $[Q,[dQ,dQ]]=0$, we obtain
(introducing the $Q^{\alpha\beta}$ and $J^{\gamma\delta}$
as the matrix element of $Q$ and $J$)
\begin{equation}
\omega=d \theta \equiv \frac{1}{2}\oint d \varphi \oint d\varphi^{\prime}
dJ^{\alpha\beta}(\varphi)\Omega_{\beta\delta;
\gamma\alpha}(\varphi, \varphi^\prime) 
dQ_{\varphi}^{\gamma\delta}(\varphi^\prime),
\label{eq:6}
\end{equation} 
where
\begin{equation}
\Omega_{\beta\delta;\gamma\alpha}(\varphi, \varphi^\prime) =
2\left(-\delta^{\beta\gamma} Q^{\delta\alpha}
+\delta^{\alpha\delta} Q^{\beta\gamma}\right)\delta(\varphi-\varphi^\prime). 
\label{eq61}
\end{equation} 
Introducing the tensor notation $(A\otimes B)_{\alpha\beta;\gamma\delta}
=A_{\alpha\gamma}B_{\beta\delta}$, we have $\Omega=
2(Q\otimes I-I\otimes Q)$. 
The notation is to be understood as including the Dirac delta-function.
Also, in order to simplify the calculation,  we  transform the original $Q\in
{\cal G}$ into $Q^2=-I/4$ which always can be
achieved by a suitable addition of constant \cite{oh}. Note that adding a
constant to $Q$ does not change the action (\ref{lagrangian}), the current
$J_\varphi^a,$ and the symplectic structure (\ref{eq61}).

In general, the presymplectic form $\Omega$ of (\ref{eq:6})  and (\ref{eq61})
is degenerate and we must calculate the Poisson bracket on the reduced phase
space. In order to achieve this, we introduce the projection operator 
$P$ and the inverse $\Omega^{-1}$ which is defined on the reduced phase space,
and  satisfy the following relations \cite{bak}
\begin{equation}
\Omega \Omega^{-1}= I\otimes I-P,~~~~ P^2=P,~~~~P\Omega=
\Omega^{-1}P
=0.
\end{equation}
Using $\Omega=2(Q\otimes I-I\otimes Q)$, we find the following solution
of the above equation
\begin{equation}
\Omega^{-1}= \frac{1}{2}(I\otimes Q-Q\otimes I), ~~~ P=\frac{1}{2}
(I\otimes I-4 Q\otimes Q).
\end{equation}
The above analysis yields the following Poisson bracket 
\begin{equation}
\{ J^{\alpha\beta}(\varphi),~ Q^{\gamma\delta} (\varphi') \}=
\Omega^{-1}_{\gamma\alpha;\beta\delta}(\varphi,\varphi')=\frac{1}{2} 
\left(\delta^{\beta\gamma} Q^{\alpha\delta}
-\delta^{\alpha\delta} Q^{\gamma\beta}\right)\delta (\varphi -\varphi '),
\label{eq:7}
\end{equation}
or in generator form
\begin{eqnarray}
 \{J^a_\varphi(\varphi ),~Q^b(\varphi ')\}=f^{abc} Q^c(\varphi)
\delta (\varphi -\varphi ').
\label{done}
\end{eqnarray}
Then, we compute the bracket $\{J^a_\varphi(\varphi),~
J^b_\varphi(\varphi^\prime )\}$. In order to properly antisymmetrize
with respect to the interchange of $(a, \varphi)\leftrightarrow(b,
\varphi^\prime)$, we calculate 
\begin{eqnarray}
 \{J^a_\varphi(\varphi),~J^b_\varphi(\varphi^\prime )\}&=&
\gamma\frac{1}{2}\left(\{J^a_\varphi(\varphi),~f^{bcd}Q^c(\varphi^\prime ) 
\partial_{\varphi^\prime}Q^d(\varphi^\prime )\}+
\{f^{acd}Q^c(\varphi) \partial_{\varphi}Q^d(\varphi ),
~J^b_{\varphi}(\varphi^\prime)\}\right)\nonumber\\
&=&
f^{abc}J^c_\varphi(\varphi )\delta (\varphi -\varphi ')+
\frac{\gamma}{2}\left(J^{ab}(\varphi) +J^{ab}(\varphi') \right)
\delta^\prime (\varphi -\varphi '),
\label{generator}
\end{eqnarray}
 where $~^\prime=\partial/\partial\varphi$ and we defined the symmetric 
tensor product 
\begin{equation}
J^{ab}(\varphi) =f^{ace}f^{bde}Q_c(\varphi )Q_d(\varphi ).
\label{eq:90}
\end{equation}
Also,  one can show, using the Jacobi identities for the structure constants,
\begin{equation}
\{J^a_\varphi(\varphi ),~J^{bc}(\varphi^\prime)\} =
\left(f^{abd}J^{dc}(\varphi) +f^{acd}J^{bd}(\varphi)\right) 
\delta (\varphi -\varphi ').
\label{eq:99}
\end{equation}
$J^a$ and $J^{bc}$ form an enlarged current algebra.
This type of current algebra ((\ref{generator}) (\ref{eq:90})
(\ref{eq:99}) ) first appeared in the analysis of
$O(N)$  nonlinear sigma model through the Dirac constraint analysis \cite{bowi}. 
Later, it was generalized to arbitrary Riemannian manifolds
with unconstrained variables using the Killing symmetry
\cite{forg}. 

\section{Diffeomorphism Algebra}

We first compute diffeomorphism (Diff) algebra directly from the  action 
(\ref{lagrangian}).
  Under diffeomorphism $\delta_f Q^a  =f^\varphi \partial_\varphi Q^a$, 
$\delta_f L=0$, and the Noether charge becomes 
\begin{eqnarray}
C(f)=\frac{\partial L}{\partial \dot{Q}^a} \delta_f Q^a 
= \gamma^{-1}\oint_{\partial D} 
d \varphi f^{\varphi} <J_{\varphi} J_{\varphi}>, 
\label{char}
\end{eqnarray}
where one used $\gamma^2 <\partial_\varphi Q\partial_\varphi Q>
=<J_{\varphi} J_{\varphi}>$.  It is more convenient to use the matrix
expression of  (\ref{generator}) from here on;
\begin{eqnarray}
\{ J^{\alpha\beta}(\varphi),~ J^{\gamma\delta} (\varphi') \}&=&
\frac{1}{2}\left(\delta^{\beta\gamma} J^{\alpha\delta}
-\delta^{\alpha\delta} J^{\gamma\beta}\right)(\varphi)
\delta (\varphi -\varphi ')
\nonumber\\
&-&\frac{\gamma}{2}\left(\frac{1}{2}
\delta^{\beta\gamma}\delta^ {\alpha\delta}
+Q^{\alpha\delta}(\varphi) Q^{\gamma\beta}(\varphi)
+Q^{\alpha\delta}(\varphi^\prime) Q^{\gamma\beta}
(\varphi^\prime)\right)\delta^\prime (\varphi -\varphi '),
\label{eq:70}
\end{eqnarray}
Then, a straight forward computation produces the Virasoro algebra
without a central charge
\begin{eqnarray}
\{ C(f), C(g) \} = C([f,g]).
\label{dewit}
\end{eqnarray}
where $[f,g]=f^\prime g-f g^\prime$.

The above derivation of the Virasoro 
started directly from the reduced action on the boundary. 
However, one can also first compute Diff charge from the bulk action
(\ref{eq:2}) by the Noether procedure, and then reduce it to a boundary Diff
charge using the solution (\ref{eq:3}). As in the pure Chern-Simons case,
the ensuing algebra depends on the boundary condition \cite{ban,ohpark}.
We  show that with a suitable choice of boundary condition,
Diff algebra becomes the Virasoro algebra with classical central term. 
Let us start with the Lagrangian (\ref{eq:2})  and consider the response of $L$
to a   spatial and time-independent diffeomorphism:
\begin{eqnarray}
\delta_f x^{\mu} &=&-\delta^{\mu}_{~ i} f^i,
~ \delta_f Q^a  =f^j \partial_j Q^a,\nonumber \\
\delta_f A^a _i &=&f^j \partial_j A^a_i +
(\partial_i f^j) A_j^a, \nonumber \\
\delta_f A^a _0 &=&f^j \partial_j A^a_0.
\label{vari}
\end{eqnarray}
We find that 
\begin{eqnarray}
\delta_f L &=& 
\frac{\kappa}{4 \pi} \int_{D} d^2 x 
\epsilon^{ij}  \partial _k \left[
 f^k \left( (A_i+2\alpha[Q,D_iQ])^a \dot{A}^a_j
+\alpha [Q, F_{ij}]^a\dot Q^a- A^a_0 G^a_{ij}\right)\right]
\nonumber \\ &=&\frac{\kappa}{4 \pi} \oint_{\partial D} 
d \varphi f^r \left[(A_r+2\alpha[Q,D_rQ])^a \dot{A}^a_{\varphi}
 -\dot{A}^a_r (A_{\varphi}+2\alpha[Q,D_\varphi Q])^a\right.
\nonumber\\
 &&~~~~~~~~~~~~~~~~~~~~~~~~~+
\left.\alpha \epsilon^{ij}[Q, F_{ij}]^a\dot Q^a
- A^a_0 \epsilon^{ij}G_{ij}^a \right].
\label{boundary}
\end{eqnarray} 
In order to have Diff  invariance, we must have $\delta_f L=\frac{d}{dt} X$,
and there are two possible boundary conditions. 

The first one is  to choose $f^r |_{\partial D}=0$ so that  $X=0$.
This boundary condition results in Diff only {\it along} the circle
($\partial D$). The Noether  charge for this Diff becomes
\begin{eqnarray}
C(f)=\frac{\partial L}{\partial \dot{A}^a_i} \delta_f A^a_i 
+\frac{\partial L}{\partial \dot{Q}^a} \delta_f Q^a
= \gamma^{-1}\oint_{\partial D} 
d \varphi f^{\varphi} <J_{\varphi} J_{\varphi}>
\end{eqnarray}
where in the second line we imposed the constraint $G^a_{ij}=0$ and its
solution (\ref{eq:3}). Note that the above equation is
the same as eq. (\ref{char}), and we again get the Virasoro algebra
without central charge.
The second boundary condition corresponds to extending the solution
(\ref{eq:3})  in the bulk to the boundary with an extra condition
$A_r=c[Q,\partial_r Q]|_{\partial D}$=constant \cite{crux}. Then, the last line
vanishes in (\ref{boundary}), and it becomes $\frac{dX}{dt}$ with 
$X=-\gamma^{-1} \oint_{\partial D} d \varphi 
f^r <J_r J_{\varphi}>,~(J_r=\gamma [Q,\partial_rQ]$=constant). 
After calculating the Noether charge, and  imposing the solution, we get
\begin{eqnarray}
C(f) &=&\frac{\partial L}{\partial \dot{A}^a_i} \delta_f A^a_i
 +\frac{\partial L}{\partial \dot{Q}^a} \delta_f Q^a
-X, \nonumber \\
&=& \gamma^{-1}
\oint _{\partial D} d \varphi 
\left( f^{\varphi}< J_{\varphi}J_{\varphi}>+ 2 f^r <J_r J_{\varphi}>\right ).
\label{cent}
\end{eqnarray}
A straight forward computation yields
 \begin{eqnarray}
\{C(f), C(g)\} &=&\gamma^{-1}
\oint _{\partial D} d \varphi 
\left[ [f^{\varphi}, g^{\varphi}]
< J_{\varphi}J_{\varphi}>+ 
2 [f^{r\prime}g^{\varphi}- f^{\varphi}g^{r\prime}]<J_r J_{\varphi}>
\right. \nonumber\\
&&\left. +2[f^{r\prime}g^r- f^rg^{r\prime}](
<J_rQJ_rQ>+\frac{1}{4}<J_rJ_r>)\right]
\label{hope}
\end{eqnarray}
In general, this bracket does not satisfy the Jacobi identity
for arbitrary $f^r$.  One  possible  choice
which satisfies the Jacobi identity is given by  
$f^r|_{\partial D}\propto \partial _{\varphi} {f^{\varphi}|_{\partial D}}$
and $g^r|_{\partial D} 
\propto \partial _{\varphi} g^{\varphi}|_{\partial D}$ 
(we choose the proportionality constant equal to 1)
along with the condition that $J_r^a=\gamma[Q, \partial_r Q]^a$=
constant.
Then, the above equation (\ref{hope}) becomes the Virasoro algebra
with classical central term;
\begin{equation}
\{ C(f), C(g) \} =C([f,g]) -
\frac{ < J_r J_r>}{\gamma} 
\oint_{\partial D} d \varphi \left( f^{\prime\prime\prime}g-
fg^{\prime\prime\prime}\right). 
\label{virasoro1}
\end{equation}
Thus, in  contrast to the Diff along 
the circle ($\partial D$),  Diff
which  deforms {\it across} the  boundary yields the Virasoro algebra
with the classical  third order derivative central term.
This is similar to what happens in the pure Chern-Simons theory 
\cite{ban,ohpark}. 

The above results also suggest that an enlarged Virasoro algebra
can be constructed.  Let us consider the expression
(\ref{cent}) directly on the boundary with some constant $J_r$. 
Then the  $<J_rQJ_rQ>$ term in (\ref{hope}) does not 
become constant, and we must choose $f^r|_{\partial D}\propto  
{f^{\varphi}|_{\partial D}}$ and $g^r|_{\partial D}  \propto
g^{\varphi}|_{\partial D}$ in order to satisfy the Jacobi identity.  This
motivates introducing  the elements of symmetric tensor product of $Q$'s
defined by
\begin{eqnarray}
S(f)&=&J_r^{\delta\alpha}J_r^{\gamma\beta}S_{\alpha\beta;\gamma\delta}(f),
\nonumber\\
S_{\alpha\beta;\gamma\delta}(f)&=&2\gamma^{-1}
\oint_{\partial D}  d \varphi f( \varphi)(Q\otimes Q)
_{\alpha\beta;\gamma\delta},
\end{eqnarray}
and  postulating the following  enlarged Virasoro algebra; 
\begin{eqnarray}
\{ C(f), C(g) \} &=&C([f,g]) +S([f,g])+
k\oint_{\partial D} d \varphi \left(
f^{\prime\prime\prime}g-fg^{\prime\prime\prime}\right), \nonumber\\ 
\{C(f),~ S(g)\}&=&- S((fg)^\prime),~~~~\{S(f),~S(g)\}=0,
\label{symmetric2}
\end{eqnarray}
where $k$=constant.
The third order derivative central charge has been put in by hand
and the above algebra satisfies the Jacobi identity. It is to be remarked that
even if the Poisson bracket does not carry the central term classically,
one can explicitly show that it arises from the standard quantum mechanical 
normal ordering \cite{godd} of $C$ and $S$. The chiral NLSM on the boundary
has a central term with only first order derivative in (\ref{hope}). 
It would be interesting to find some explicit model in which the enlarged
Virasoro algebra (\ref{symmetric2}) can be realized classically.

\section{Conclusion}
We  have constructed a
boundary model in which the Chern-Simons gauge field and NLSM have bulk
topological interaction with a manifest gauge invariance. 
We showed that the modified Gauss's law can be solved explicitly when the 
NLSM is defined on the Hermitian symmetric space, and  the original bulk
theory reduces to a boundary NLSM which has a time-independent 
diffeomorphism invariance.
We find the coadjoint orbit method for NLSM particularly convenient
in the process. 
We also calculated the symplectic structure and the diffeomorphism
algebra. If we first calculate the Diff charge directly from the bulk
action by the Noether procedure, and then reduce it to a boundary
Diff charge using the solution of Gauss's law constraint,  the ensuing  Diff
algebra which  corresponds to deformation {\it across} the  boundary yields the
Virasoro algebra with {\it classical} central term. We also discovered an
enlarged Virasoro algebra with elements of symmetric tensor product of $Q$'s.
It would be interesting to study the enlarged Virasoro algebra in detail, 
and to investigate whether the present result can find some application in the
black hole physics.   

\section{Acknowledgements}
I like to thank Joohan Lee for useful discussions, and Prof. 
R. Jackiw for his kind hospitality at CTP of MIT where part of this work was done. 
This work is supported by the Korea Research 
Foundation through program 1998-015-D00034.


\begin{references}
\bibitem{crux} This condition forces the variables $Q^a$'s to a subset
of all possible configurations on the boundary. A more detailed study 
will be addressed elsewhere.
\end{references}
\end{document}